\documentclass[english]{article}
\usepackage[T1]{fontenc}
\usepackage[latin9]{inputenc}
\usepackage{color}
\usepackage{amsmath}
\usepackage{graphicx}
\usepackage{setspace}

\DeclareRobustCommand{\greektext}{%
  \fontencoding{LGR}\selectfont\def\encodingdefault{LGR}}
\DeclareRobustCommand{\textgreek}[1]{\leavevmode{\greektext #1}}
\DeclareFontEncoding{LGR}{}{}

\usepackage{babel}

\begin{document}

\title{\textcolor{black}{Can the Tajmar effect be explained using a modification
of inertia?}}

\author{\textcolor{black}{M.E. McCulloch}%
\thanks{\textcolor{black}{School of Physics, University of Exeter, Devon,
U.K, M.E.McCulloch@exeter.ac.uk}%
}\textcolor{black}{\ }%
\thanks{\textcolor{black}{University of Plymouth, Devon, UK, mike.mcculloch@plymouth.ac.uk}%
}}
\maketitle
\begin{abstract}
\begin{singlespace}
\noindent \textcolor{black}{The Tajmar effect is an unexplained acceleration
observed by accelerometers and laser gyroscopes close to rotating
supercooled rings. The observed ratio between the gyroscope and ring
accelerations was $3\pm1.2\times10^{-8}$. Here, a new model for inertia
which has been tested quite successfully on the Pioneer and flyby
anomalies is applied to this problem. The model assumes that the inertia
of the gyroscope is caused by Unruh radiation that appears as the
ring and the fixed stars accelerate relative to it, and that this
radiation is subject to a Hubble-scale Casimir effect. The model predicts
that the sudden acceleration of the nearby ring causes a slight increase
in the inertial mass of the gyroscope, and, to conserve momentum in
the reference frame of the spinning Earth, the gyroscope rotates clockwise
with an acceleration ratio of $1.78\pm0.25\times10^{-8}$ in agreement
with the observed ratio. However, this model does not explain the
parity violation seen in some of the gyroscope data. To test these
ideas the Tajmar experiment (setup B) could be exactly reproduced
in the southern hemisphere, since the model predicts that the anomalous
acceleration should then be anticlockwise.}\end{singlespace}

\end{abstract}
\textcolor{black}{\pagebreak{}}

\section{\textcolor{black}{Introduction}}

\textcolor{black}{The Tajmar effect is a small unexplained acceleration
observed in accelerometers and laser gyroscopes close to supercooled
rotationally-accelerated rings of niobium, aluminium, stainless steel
and other materials {[}1,2,3{]}. The effect is similar to the Lense-Thirring
effect (frame-dragging) predicted by General Relativity, but is 20
orders of magnitude larger. The effect has not yet been reproduced
in another laboratory.}

\textcolor{black}{{[}4{]} proposed an explanation for the anomaly
that relies on a Higgs mechanism that causes the graviton to gain
mass. This theory was called the gravitometric London effect, but
it seems to have been discredited because the inception of the Tajmar
effect does not coincide with the superconducting transition temperature,
only with very low temperatures.}

\textcolor{black}{The model suggested here as an explanation for the
effect was proposed by {[}5{]}. It assumes that the inertial mass
of an object with respect to an attracting body is caused by Unruh
radiation which is generated by the relative acceleration of the two
bodies, and that this radiation is subject to a Hubble-scale Casimir
effect (in which longer waves are increasingly disallowed). The model
could be called Modified Inertia due to a Hubble-scale Casimir effect
(MiHsC) or perhaps Quantised Inertia, and was tested by {[}5{]} on
the Pioneer anomaly (observed by {[}6{]}). In {[}5{]} the inertial
mass ($m_{I}$) was defined as}

\textcolor{black}{\begin{equation}
m_{I}=m_{g}\left(1-\frac{\beta\pi^{2}c^{2}}{\left|a\right|\Theta}\right)\end{equation}
}

\begin{singlespace}
\noindent \textcolor{black}{where $m_{g}$ is the gravitational mass,
$\beta=0.2$ (empirically derived by Wien as part of Wien's law),
c is the speed of light, $\Theta$ is the Hubble diameter ($2.7\times10^{26}m$,
derived from {[}10{]} and the magnitude of the acceleration (a) in
this case was the acceleration of the Pioneer craft relative to their
main attractor: the Sun. This model predicted a small loss of inertial
mass for the Pioneer spacecraft that increased their Sunward acceleration
by an amount close to the observed Pioneer anomaly (when the spacecraft
were unbound beyond 10AU from the Sun).}
\end{singlespace}

\noindent \textcolor{black}{Reference {[}7{]} applied MiHsC to the
unexplained velocity jumps observed in Earth flybys of interplanetary
probes (the flyby anomalies observed by {[}8{]}) and found that these
anomalies could be reproduced quite well if the acceleration in equation
1 was taken to be that of the spacecraft relative to }\textbf{\textcolor{black}{all}}\textcolor{black}{
the particles of matter in the spinning Earth. Using MiHsC and the
conservation of momentum the predicted anomalous jump for spacecraft
passing by the spinning Earth was}

\textcolor{black}{\begin{equation}
dv'=\frac{\beta\pi^{2}c^{2}}{\Theta}\left(\frac{v_{2}}{\left|a_{2}\right|}-\frac{v_{1}}{\left|a_{1}\right|}\right)\end{equation}
}

\begin{singlespace}
\noindent \textcolor{black}{where $a_{1}$ and $a_{2}$ were the average
accelerations of all the matter in the spinning Earth seen from the
point of view of the incoming ($a_{1}$) and outgoing ($a_{2}$) craft.
This formula predicted the observed flyby anomalies quite well and
was similar to the empirical formula suggested by {[}8{]}.}

\noindent \textcolor{black}{The Tajmar effect is similar to the flyby
anomalies, although instead of being an anomalous acceleration of
a spacecraft close to a spinning planet, it is an anomalous acceleration
observed in laser gyroscopes close to a super-cooled spinning ring.
Therefore, in this paper MiHsC is applied to the Tajmar effect, but
also, following Mach's principle, in this paper the relative accelerations
of the fixed stars are also considered.}
\end{singlespace}

\section{\textcolor{black}{Method}}

\textcolor{black}{The assumed experimental set up is that of {[}3{]}
and their set-up configuration B, and is shown in Figure 1, with a
rotating super-cooled ring of radius r. A laser gyroscope of mass
m is located symmetrically above the ring. The fixed stars are also
shown schematically. They have a huge combined mass, but are very
far away. Assuming the conservation of the angular momentum of the
gyroscope (subscript G) from one time to another (subscripts 1 and
2) in the reference frame of the fixed stars (subscript s) we have}

\textcolor{black}{\begin{equation}
I_{G1}w_{Gs1}=I_{G2}w_{Gs2}\end{equation}
}

\noindent \textcolor{black}{where I is the moment of inertia ($I=\Sigma mr^{2}$)
and w is the angular momentum ($w=v/r$). For an segment of the circular
laser gyroscope (see 'g' on Figure 1) this simplifies to the conservation
of momentum parallel to the ring}

\textcolor{black}{\begin{equation}
m_{g1}v_{gs1}=m_{g2}v_{gs2}\end{equation}
}

\textcolor{black}{Replacing the inertial masses with the modified
inertia of {[}5{]} leaves}

\textcolor{black}{\begin{equation}
v_{gs1}\left(1-\frac{\beta\pi^{^{2}}c^{2}}{\left|a_{g1}\right|\Theta}\right)=v_{gs2}\left(1-\frac{\beta\pi^{^{2}}c^{2}}{\left|a_{g2}\right|\Theta}\right)\end{equation}
}

\textcolor{black}{and rearranging}

\textcolor{black}{\begin{equation}
v_{gs2}-v_{gs1}=\frac{\beta\pi^{^{2}}c^{2}}{\Theta}\left(\frac{v_{gs2}}{\left|a_{g2}\right|}-\frac{v_{gs1}}{\left|a_{g1}\right|}\right)\end{equation}
}

\noindent \textcolor{black}{Of course, this is similar to equation
2, which was derived from MiHsC for the flyby anomalies. For this
new case, the initial and final accelerations ($a_{g1}$ and $a_{g2}$)
of the gyroscope with respect to all the surrounding masses now need
to be defined. First we assume that because of cooling the temperature-dependent
acceleration of nearby atoms is small. We can say that the acceleration
relative to the atoms in the Earth is zero since the experiment is
solidly fixed to the Earth. So, before the ring accelerates the gyroscope
sees only an acceleration of the fixed stars since it is on the spinning
Earth. These are far away, but their combined mass is huge. The rotational
acceleration with respect to the fixed stars ($a_{s}$) of an object
fixed to the Earth at the latitude of Seibersdorf in Austria where
the experiment was performed (at 48$^{\text{0}}N$) is the same as
the Coriolis acceleration: fv, where $f\sim0.0001s^{-1}$ in mid-latitudes,
and v, the spin velocity of the Earth at this latitude is 311 m/s,
so $a_{s}=0.0311m/s^{2}$. To this should be added the acceleration
due to the Earth's orbit around the Sun (so we have: $a_{s}=(0.0311+0.006)m/s^{2})$.
The acceleration due to the Sun's orbit around the galaxy is far smaller
and can be neglected. So in the above formula $a_{g1}=0.0371m/s^{2}$.}

\noindent \textcolor{black}{The sudden acceleration of the Tajmar
ring causes an acceleration of $a_{R}=33\, rad/s^{2}=2.5\, m/s^{2}$
(since the radial position of the gyroscope was 0.075 m). Therefore
$a_{g2}=fn(a_{s},a_{R})$. However, to find the average acceleration
we have to consider the relative importance of the fixed stars and
the ring for determining the inertia of the gyroscope. We will assume
here that the importance of an object for the inertia of another one
is equivalent to its gravitational importance, which is proportional
to its mass over the distance squared. This details of this assumption
do not effect the final result as we will see. Therefore $a_{g2}$
is}

\textcolor{black}{\begin{equation}
a_{g2}=\frac{\frac{m_{s}}{r_{s}^{2}}a_{s}+\frac{m_{R}}{r_{R}^{2}}a_{R}}{\frac{m_{s}}{r_{s}^{2}}+\frac{m_{R}}{r_{R}^{2}}}\end{equation}
}

\noindent \textcolor{black}{where $m_{s}$ is the mass of all the
fixed stars, and $r_{s}$ is their mean distance away and $m_{R}$
is the mass of the ring and $r_{R}$ is its distance away. Using equation
7 in equation 6 gives}

\textcolor{black}{\begin{equation}
v_{gs2}-v_{gs1}=\frac{\beta\pi^{2}c^{2}}{\Theta}\left(\frac{v_{gs2}}{\left|\frac{\frac{m_{s}}{r_{s}^{2}}a_{s}+\frac{m_{R}}{r_{R}^{2}}a_{R}}{\frac{m_{s}}{r_{s}^{2}}+\frac{m_{R}}{r_{R}^{2}}}\right|}-\frac{v_{gs1}}{\left|a_{s}\right|}\right)\end{equation}
}

\begin{singlespace}
\noindent \begin{flushleft}
\textcolor{black}{These values were approximated as a total stellar
mass of $m_{s}\sim2.4\times10^{52}kg$ from {[}9{]}, at a distance
of $r_{s}\sim2.7\times10^{26}m$ ($r_{s}=2c/H$, derived from the
Hubble constant, H, from {[}10{]}), and a ring mass of $m_{R}\sim0.336kg$
(stainless steel has a density of about 8000 kg/m$^{\text{3}}$, and
the ring had a circumference of 2{*}\textgreek{p}{*}0.075, a height
of 0.015 m and a width of 0.006 m) and a ring distance of $r_{R}\sim0.039m$
(the distance from the above-ring gyroscope to the nearest part of
the ring). Using these values we have}
\par\end{flushleft}
\end{singlespace}

\textcolor{black}{\begin{equation}
v_{gs2}-v_{gs1}=\frac{\beta\pi^{2}c^{2}}{\Theta}\left(\frac{v_{gs2}}{\left|\frac{0.96a_{s}+220a_{R}}{220}\right|}-\frac{v_{gs1}}{\left|a_{s}\right|}\right)\end{equation}
}

\textcolor{black}{We can therefore neglect $a_{s}$ in the denominator
of the first term in brackets (this simplification similarily applies
to the aluminium and niobium rings) to give}

\textcolor{black}{\begin{equation}
v_{gs2}-v_{gs1}\sim\frac{\beta\pi^{2}c^{2}}{\Theta}\left(\frac{v_{gs2}}{\left|a_{R}\right|}-\frac{v_{gs1}}{\left|a_{S}\right|}\right)\end{equation}
}

\textcolor{black}{Since $v_{gs2}\sim v_{gs1}$and $a_{R}$ is 2 orders
of magnitude greater than $a_{s}$ we can say}

\textcolor{black}{\begin{equation}
dv'\sim-\frac{\beta\pi^{2}c^{2}}{\Theta}\frac{v_{gs1}}{\left|a_{s}\right|}\end{equation}
}

\textcolor{black}{Differentiating to find the resulting acceleration}

\textcolor{black}{\begin{equation}
da'\sim-\frac{\beta\pi^{2}c^{2}}{\Theta}\frac{a_{gs}}{\left|a_{s}\right|}\sim\frac{-6.6\times10^{-10}}{0.0371}a_{gs}\sim-1.78\pm0.25\times10^{-8}a_{gs}\end{equation}
}

\textcolor{black}{Since $a_{gs}$(the rotational acceleration of the
gyroscope with respect to the fixed stars) is anticlockwise in the
northern hemisphere because of the Earth's spin, equation 12 implies
that the anomalous rotational acceleration of the gyroscope will be
clockwise, as observed by {[}3{]} for set-up B. The error of 0.25
was derived by assuming a 9\% error in the Hubble constant (and therefore
the estimated Hubble diameter) following {[}10{]}. The predicted acceleration
is close to the ratio of $da'/a_{ar}$ that was observed by {[}3{]}.
Using niobium, aluminium and stainless steel rings they observed a
velocity coupling between the rings and the gyroscope (equal to the
acceleration coupling factor since acceleration here is angular velocity
times a constant radius) which increased from zero above the transition
temperature at 25K, to reach values of about$-3\pm1.2\times10^{-8}$
at 5K (see {[}3{]}, figures 3 and 6). This ratio agrees within error
bars with the ratio predicted by equation 12, but the parity violation
({[}3{]}, figure 2), only observed for their setup A, is not predicted.
Rings of YBCO and Teflon produced similar results.}

\section{\textcolor{black}{Discussion}}

\textcolor{black}{In the model used here (MiHsC) the inertial mass
of the gyroscope is assumed to be determined by Unruh radiation that
appears as it accelerates with respect to every other mass in the
universe. The Unruh radiation is also subject to a Hubble-scale Casimir
effect. Before its surroundings are cooled the gyroscope sees large
accelerations due to the vibration of nearby atoms, it is surrounded
by Unruh radiation of short wavelengths, MiHsC has only a small effect,
and the inertial mass of the gyroscope is close to its gravitational
mass. The nearby atomic accelerations reduce when the surroundings
(the cryostat) are cooled, so that the inertia of the gyroscope is
more sensitive to the accelerations of the fixed stars (it is on the
rotating Earth). This is a small acceleration, so the Unruh waves
it sees are long and a greater proportion are disallowed by MiHsC,
so the gyroscope's inertial mass (from equation 1) falls very slightly
below its gravitational mass. In this case it looses $2\times10^{-10}kg$
for every kilogram of mass. However, when the nearby ring accelerates,
the gyroscope suddenly sees the higher accelerations of the ring,
the Unruh waves shorten, fewer are disallowed, and its inertial mass
increases again following equation 1. The important point is that
to conserve momentum the gyroscope must accelerate in the opposite
direction to the Earth's spin: clockwise.}

\textcolor{black}{This can be summarised by saying that if an object
A changes its inertial mass by MiHsC because of changes in the relative
acceleration of an object B, then A must accelerate to conserve the
momentum of the joint system (A+B) in the reference frame of the fixed
stars.}

\textcolor{black}{A more accurate calculation of the predicted acceleration
ratio using equation 8 gives $-1.8x10^{-8}$ and only a slow decay
with distance $r_{R}$. If the velocity jump derived from equation
8 is calculated for gyroscopes at various distances from the ring
($r_{R}$) starting at $r_{R}=0.016m$ and increasing, then the new
MiHsC effect decreases to half its original size at $r_{R}=56m$,
ie: 56 m away. However, this slow decrease only applies to a uniformly
cold environment such as space, and does not take account of the warmer
environment further away from the ring whose thermal accelerations
would hide the effect of MiHsC. As discussed above (Equation 7) the
importance of the cold ring for the inertia of the accelerometer decreases
inversely as the square of the distance, so, for a gyroscope, say,
four times as far from the cold environment surrounding the ring as
the above-ring gyroscope, the MiHsC effect due to the cold ring will
be 16 times smaller, and far less detectable. It is interesting in
this respect, that while the gyroscope was exposed on the outside
to the cold liquid helium the anomalous signal did not decay with
distance ({[}3{]}, setup A) whereas when it was moved outside the
cryostat the signal did decay quickly ({[}3{]}, setup C).}

\textcolor{black}{One criticism of MiHsC is that it makes no mention
of the accelerations of the atoms in spinning stars, for example.
Why should the acceleration of the accelerometer with respect to the
fixed stars be only $0.0371m/s^{2}$ as assumed here, when the fixed
stars are rotating with larger accelerations than this and contain
atoms with very large accelerations?}

\section{\textcolor{black}{Tests}}

\textcolor{black}{One way to test these ideas would be to reproduce
setup B of the Tajmar experiment (see {[}3{]}) in the southern hemisphere,
since MiHsC predicts that then the anomalous acceleration of the gyroscope
should be anticlockwise instead of clockwise. This hemisphere-dependence
may already have been observed by {[}2{]}, who mentioned that the
Earth's rotation may therefore be somehow involved.}

\textcolor{black}{The size of $a_{s}$ in equation 11 could be reduced
by performing the Tajmar experiment in a frame that rotates to follow
the fixed stars, to counter the Earth's rotation. A partial example
of such a system would be a Foucault pendulum. The anomalous jump
seen should then be larger. This test would be easier to achieve at
the Earth's poles.}

\textcolor{black}{The Tajmar experiment could be repeated, slowing
the spin of the ring so that the ring's acceleration relative to the
gyroscope is close to that of the fixed stars ($0.036\, m/s^{2}$)
then according to equation 10 the anomalous jump should dissapear.
For even lower accelerations (values of $a_{R}$) the anomalous effect
should reverse. Equation 9 implies that for very small values of $a_{R}$
the anomalous acceleration is predicted to be a maximum of about 229
times greater than the anomaly predicted by equation 11. However,
at these lower accelerations the acceleration ratio could be undetectable
given the precision of the gyroscopes.}

\textcolor{black}{Since the relative acceleration of the fixed stars
has a different value depending on latitude, the inertial mass of
objects, especially supercooled objects, should vary slightly with
latitude. This should be apparent as a change in weight when objects
are moved latitudinally. A move towards the equator would decrease
the apparent acceleration of the fixed stars, so according to MiHsC
the inertial mass should decrease, the object would respond more actively
to the Earth's gravity (ie: it would not follow a straight line through
space so dutifully) and so its gravitational mass would appear to
increase. In a move towards the pole, the weight would seem to decrease.}

\section{\textcolor{black}{Conclusions}}

\textcolor{black}{The anomalous clockwise accelerations observed by
laser gyroscopes close to rotating super-cooled rings (see {[}3{]},
for set-up B) can be predicted by a theory (MiHsC) that assumes that
the inertial mass of the gyroscope is caused by Unruh radiation that
appears because of its mutual acceleration with the fixed stars and
the spinning ring, and that this radiation is subject to a Hubble-scale
Casimir effect (MiHsC has already been shown to offer a reasonably
successful explanation for the Pioneer and flyby anomalies.)}

\textcolor{black}{MiHsC does not explain the parity violations seen
in the results of {[}3{]} using their set-up A, in which the gyroscopes
were positioned assymetrically with respect to the spinning ring.}

\textcolor{black}{This explanation for the Tajmar effect could be
tested by exactly reproducing the experiment of {[}3{]} (their setup
B) in the southern hemisphere, since MiHsC predicts that then the
anomalous acceleration of the gyroscopes should be anti-clockwise.}

\section*{\textcolor{black}{Acknowledgements}}

\textcolor{black}{Many thanks to M.Tajmar, M.Bate, J.R.Sambles, B.Kim
and a reviewer for good advice.}

\section*{\textcolor{black}{References}}

\textcolor{black}{{[}1{]} Tajmar, M., F. Plesescu, B. Seifert, and
K. Marhold, 2007. Measurement of gravitomagnetic and acceleration
fields around rotating superconductors. Proceedings of the STAIF-2007
Conference, AIP Conference Proceedings, Vol. 880, pp. 1071-1082.}

\textcolor{black}{{[}2{]} Tajmar, M., F. Plesescu, B. Seifert, R.
Schnitzer, and I. Vasiljevich, 2008. Investigation of frame-dragging-like
signals from spinning superconductors sing laser gyrocopes. AIP conference
Proceedings, Vol. 69, pp 1080-1090.}

\textcolor{black}{{[}3{]} Tajmar, M., F. Plesescu and B. Seifert,
2009. Anomalous fiber optic gyroscope signals observed above spinning
rings at low temperature. J. Phys. Conf. Ser., 150, 032101, 2009.}

\textcolor{black}{{[}4{]} de Matos, C.J. and M. Tajmar, 2005. Gravitometric
London moment and the graviton mass inside a superconductor, Physica
C, 432, 167-172.}

\textcolor{black}{{[}5{]} McCulloch, M.E., 2007. Modelling the Pioneer
anomaly as modified inertia. }\textit{\textcolor{black}{MNRAS}}\textcolor{black}{,
376, 338-342.}

\textcolor{black}{{[}6{]} Anderson, J.D, P.A. Laing, E.L. Lau, A.S.
Liu, M.M. Nieto and S.G. Turyshev, 1998. Phys. Rev. Lett., 81, 2858.}

\textcolor{black}{{[}7{]} McCulloch, M.E., 2008. Modelling the flyby
anomalies using a modification of inertia. }\textit{\textcolor{black}{MNRAS-letters}}\textcolor{black}{,
389, L57-60.}

\textcolor{black}{{[}8{]} Anderson, J.D., Campbell, J.K., Ekelund,
J.E., Ellis J., Jordan J.F., 2008. Anomalous orbital energy changes
observed during spacecraft flybys of Earth. }\textit{\textcolor{black}{Phys.
Rev. Lett.}}\textcolor{black}{, 100, 091102.}

\textcolor{black}{{[}9{]} Funkhouser, S., 2006. The large number coincidence:
the cosmic coincidence and the critical acceleration. Proc. R. Soc
A., vol 462, no 2076, p3657-3661.}

\textcolor{black}{{[}10{]} Freedman, W.L., 2001. Final results fom
the Hubble space telescope key project to measure the Hubble constant.
ApJ, 553, 47-72.}

\textcolor{black}{\pagebreak{}}

\section*{\textcolor{black}{Figures}}

\textcolor{black}{\includegraphics{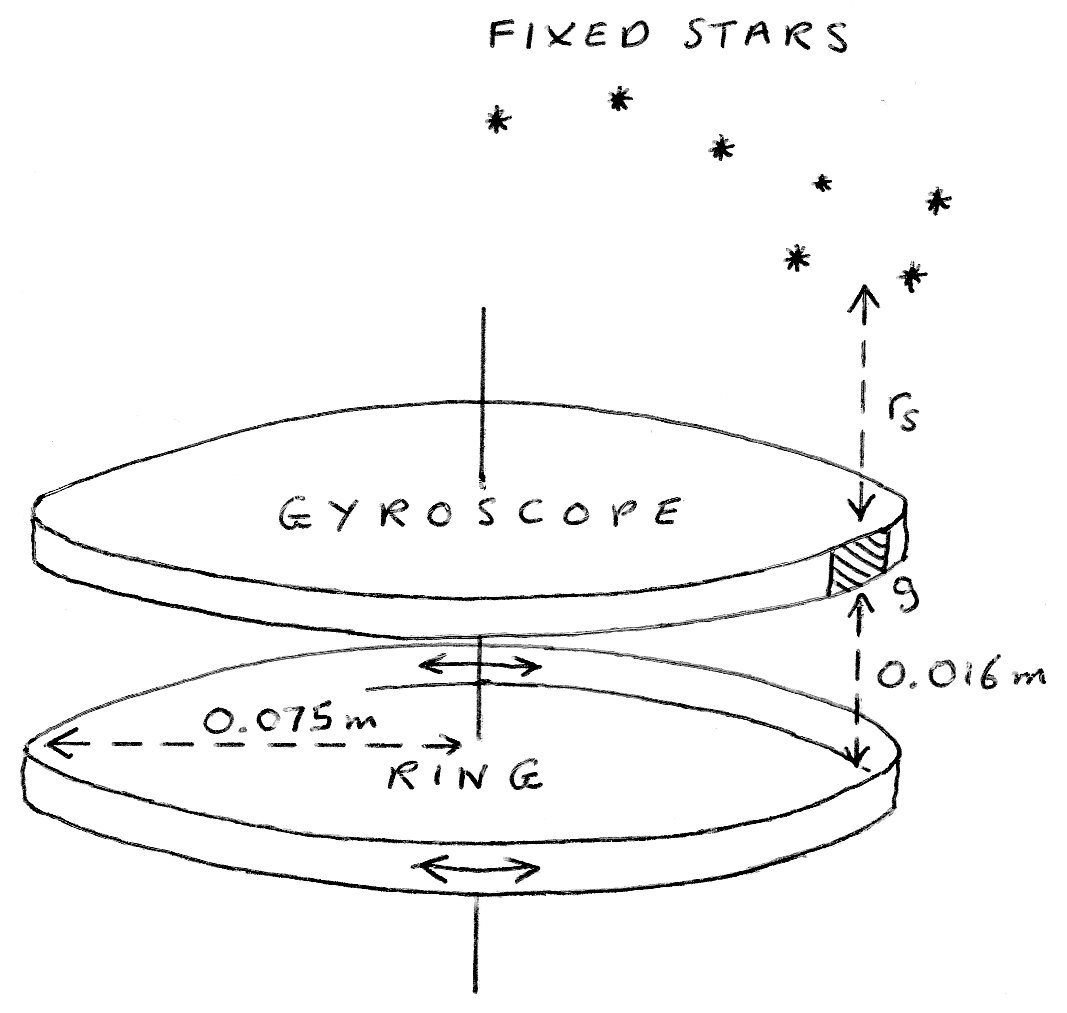}}

\textcolor{black}{Figure 1. Schematic showing the experiment of {[}3{]},
setup B. The ring, the laser gyroscope, some dimensions and the fixed
stars are shown.}
\end{document}